\documentclass[twocolumn,twocolappendix,trackchanges,resetfootnote]{aastex701}

\usepackage{CJKutf8}

\begin{document}
\begin{CJK*}{UTF8}{gbsn}

\title{Down-bending Breaks in Galactic Disks Are an Intrinsic Byproduct of Inside-out Growth}

\correspondingauthor{Min Du}
\email{dumin@xmu.edu.cn}

%\author{Liufei Chen}
\author{Liufei Chen (陈柳霏)}

\affiliation{School of Physics and Astronomy, China West Normal University, 1 ShiDa Road, Nanchong 637002, China}
\affiliation{Department of Astronomy, Xiamen University, Xiamen, Fujian 361005, China}
\email{??}

%\author{Min Du (杜敏); \href{mailto:dumin@xmu.edu.cn}{dumin@xmu.edu.cn}}
\author{Min Du (杜敏)} 
%\affiliation{Department of Astronomy, Xiamen University, Xiamen, Fujian 361005, China; \href{mailto:dumin@xmu.edu.cn}{dumin@xmu.edu.cn}}
\affiliation{Department of Astronomy, Xiamen University, Xiamen, Fujian 361005, China}
\email{dumin@xmu.edu.cn}

\author{Shuai Lu (卢帅)}
\affiliation{Department of Astronomy, Xiamen University, Xiamen, Fujian 361005, China}
\email{??}

\author{Jing Li (李静)}
\affiliation{School of Physics and Astronomy, China West Normal University, 1 ShiDa Road, Nanchong 637002, China}
\email{??}

\author{Luis C. Ho}
\affiliation{Kavli Institute for Astronomy and Astrophysics, Peking University, Beijing 100871, China}
\affiliation{Department of Astronomy, School of Physics, Peking University, Beijing 100871, China}
\email{??}

\begin{abstract}
The exponential profile has long been hypothesized as the fundamental morphology of galactic disks. The IllustrisTNG simulations successfully reproduce diverse surface-density profiles: Type~I (single exponential), Type~II (down-bending), and Type~III (up-bending), matching reasonably well with observed mass-size relations and kinematics. Type~II disks dominate the stellar mass regime $M_\star < 10^{10.6} M_\odot$ with a prevalence of $\sim$40\%, exhibiting systematically extended morphologies. Conversely, Type~III and Type~I galaxies occupy more compact configurations, following the same mass-size scaling relation. Evolutionary history analysis shows that Type~II galaxies experience minimal external perturbation, representing the intrinsic disk form—challenging conventional single-exponential paradigms. We demonstrate that Type~II breaks arise naturally via inside-out growth during $z<1$, governed by synchronized cold-gas accretion and localized specific star formation rate peaks. This mechanism also produces the characteristic U-shaped age profiles of Type~II disks. Crucially, dynamical redistribution of stars plays an unimportant role in their formation.

\end{abstract}


\section{Introduction} 

It has long been recognized that galactic disk radial surface-brightness profiles are described by an exponential function \citep{freeman1970}. However, deviations are frequently observed in the outer regions, which are typically classified into three distinct categories: Type~I (single-exponential), Type~II (down-bending), and Type~III (up-bending) \citep{erwin2005, pohlen2006}. Single-exponential (Type~I) disks are widely regarded as the natural baseline configuration in theoretical models \citep[e.g.,][]{mo1998, vandenBosch2001, dutton2009} as well as morphological decomposition in observations \citep[e.g.,][]{peng2010, diaz2016}. \citet{wangenci2022} suggested that the exponential profile of disk galaxies can be a natural product due to the viscosity of gas.

The distribution of these disk profiles exhibits a dependence on stellar mass, Hubble type, and local environment. Type~II profiles are prevalent in the mass range $\log(M_\star/M_\odot)\approx 9.5$--10.5, whereas Type~III profiles dominate the high-mass regime, $\log(M_\star/M_\odot)\gtrsim 10.5$ \citep{tang2020}. This mass dependence is reflected in morphology: Type~II profiles occur more frequently in late-type, disk-dominated galaxies, while Type~III profiles are more characteristic of early-type systems \citep{pohlen2006, gutierrez2011, laine2016, tang2020}. The Milky Way has also been found to follow a Type~II break profile \citep{wang2018mapping, lian2024, lian2025}. U-shaped average age profiles have been found to be prevalent in Type~II disk galaxies \citep{de2007,radburn2012,dale2015}, although \citet{ruiz2015} argued that the U-shaped age profiles and Type~II breaks are not necessarily coupled. Moreover, the characteristic Type~II break radius scales positively with galaxy mass and has been observed to shift outward over cosmic time \citep{trujillo2005, azzollini2008, pohlen2006}.

Various physical processes have been invoked to explain the formation of Type~III profiles, involving either internal dynamical evolution or external environmental effects. Internally, galactic bars can scatter stars from circular orbits onto large, elongated trajectories \citep{brunetti2011}, effectively transforming Type~I or~II profiles into Type~III configurations \citep{herpich2017, ruiz2017}. Externally, minor mergers and tidal interactions redistribute stellar material via tidal torques that propel inner stars outward \citep{younger2007}. Additionally, stellar material stripped from satellite galaxies during mergers adds directly to the light of the outer regions \citep{cooper2013}.

The formation mechanisms of Type~II breaks also remain a subject of active debate, primarily centered on two proposed scenarios. The first attributes Type~II formation to radial migration and resonance-driven redistribution, where non-axisymmetric features, such as bars and spiral arms, are able to transport stars from the inner disk to beyond the break radius \citep{debattista2006, herpich2017}. While this framework can explain observed U-shaped age profiles \citep{zheng2015, dale2015}, recent work suggests that such redistribution may eventually smooth out or erase these breaks \citep{sanchez2009, tang2020}.
The second scenario involves an abrupt change in the radial star formation profile. Simulation studies indicate that Type~II disks may predominantly emerge from truncated star-forming disks \citep{pfeffer2022}. Observationally, these galaxies exhibit down-bending radial star formation rate (SFR) profiles \citep{tang2020}. However, while this may explain the observed radial truncation, it struggles to account for the old stellar populations at large radii; these may generally require supplementary radial migration \citep{roskar2008, tang2020} or an extended stellar halo \citep{pfeffer2022}.

The occurrence of Type~II profiles may also be influenced by environment, with a notable suppression observed in cluster galaxies \citep{erwin2012, mondelin2025cluster}. \citet{pfeffer2022} linked this reduced Type~II fraction to the generally lower star-formation rates in cluster environments, suggesting that Type~II disks may fade into Type~I profiles once star formation ceases. This view is supported by \citet{sanchez2023}, who found a higher fraction of perturbations in field Type~II disk galaxies ($42\% \pm 19\%$) compared to Type~I ($17\% \pm 17\%$), implying that perturbations may drive Type~II formation by enhancing star formation. Conversely, \citet{mondelin2025simulation} propose that Type~II profiles originate from internal processes such as disk instabilities and resonances during early formation phases, and are subsequently eroded by the cluster environment. These conflicting perspectives highlight a key uncertainty: whether the Type~II profile is a transient feature associated with star formation or instead arises from internal disk evolution. The latter scenario would potentially challenge the traditional view of the Type~I profile as the fundamental baseline.

In this study, we analyze central disk galaxies (stellar mass $M_\star > 10^{10} M_\odot$) from the IllustrisTNG hydro-cosmological simulations, classifying systems into Types~I, II, and III through fits to their stellar mass surface density profiles. This approach directly addresses the origin of observed disk morphological diversity. We subsequently focus on deciphering Type~II break formation mechanisms and their signature U-shaped stellar age profiles. A new channel for forming Type~II break is proposed. The paper is structured as follows: Section~\ref{sec:2} details the simulation data, galaxy selection criteria, and profile extraction methodology; Section~\ref{sec:3} presents statistical properties and comparative analysis of all three disk types; Section~\ref{sec:4} investigates the physical drivers of Type~II breaks and associated star-formation trends; Section~\ref{sec:5} synthesizes implications and summarizes key conclusions.

%fig1
\begin{figure*}[!t] 
\centering
\includegraphics[width=0.88\textwidth]{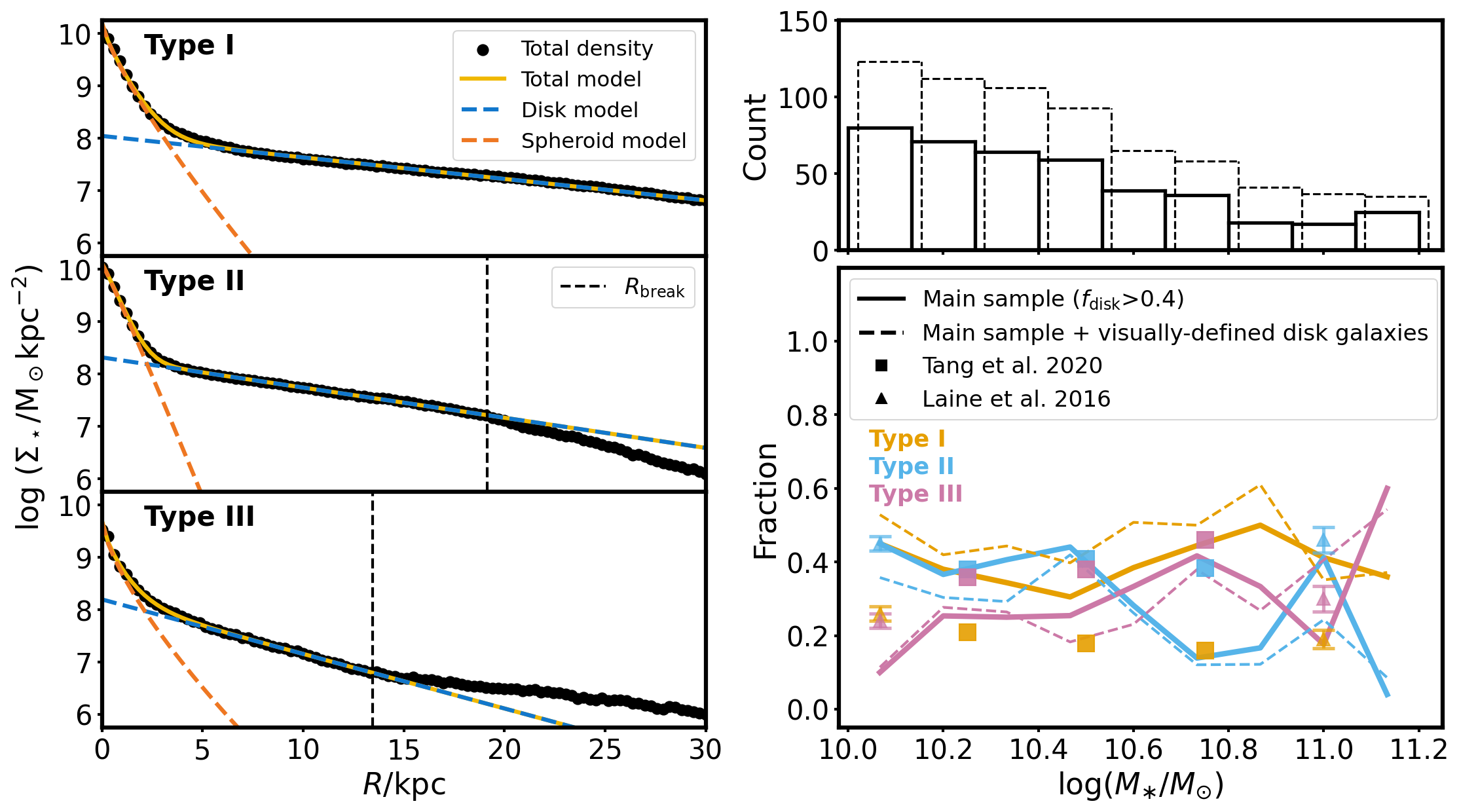}
\caption{
\textbf{Left panels:} Examples of surface density profiles (black dots) of galaxies with Type~I, II, and III disks, from top to bottom. The exponential (blue-dashed lines) and S\'ersic (orange-dashed) functions are used to decompose the disk and bulge component in morphology, their combined fitting results correspond to the yellow solid profiles. Vertical black dashed lines mark the break radius ($R_{\rm break}$), where the profile first deviates from the inner exponential part. 
\textbf{Right panels:} The fraction of galaxies with different disk types as a function of their stellar masses. Solid and dashed lines denote results from two galaxy samples: our kinematically-selected primary sample ($f_{\rm disk} > 0.4$; \citet{du2019, du2020}) and a morphologically-defined sample incorporating visually classified disks to enable direct comparison with observational surveys \citep{tang2020,laine2016}. The upper panel gives their number counts in each stellar mass bin.}
\label{fig:1}
\end{figure*}

\section{Methodology: sample selection and classification}\label{sec:2}
\subsection{The IllustrisTNG project}
The IllustrisTNG project \citep{marinacci2018, naiman2018, nelson2018, nelson2019, pillepich2018, pillepich2019, springel2018} is a suite of cosmological magnetohydrodynamical simulations executed with the moving-mesh code AREPO \citep{springel2010}, performed within the standard $\Lambda$CDM cosmology adopting \citet{Planck2016} parameters: $\Omega_m = 0.3089$, $\Omega_\Lambda = 0.6911$, $\Omega_b = 0.0486$, $h = 0.6774$, $\sigma_8 = 0.8159$, and $n_s = 0.9667$. The project includes three simulation runs (TNG300, TNG100, TNG50) with varying volumes and resolutions \citep{pillepich2018,nelson2019b}. Compared to the original Illustris simulation, TNG incorporates improved models for chemical enrichment, stellar and black hole feedback, and cosmic magnetism \citep{Weinberger2017, pillepich2018}. We utilize TNG50-1 (hereafter TNG50), which provides the highest resolution for studying stellar structures in galaxy peripheries. TNG50 employs $2 \times 2160^3$ initial resolution elements within a $\sim$50 comoving Mpc periodic volume, achieving baryon mass resolution of $8.41 \times 10^4 \,\mathrm{M_\odot}$ and dark matter particle mass of $4.57 \times 10^5 \,\mathrm{M_\odot}$. The gravitational softening length is 0.29 kpc for dark matter and stellar particles at $z = 0$, with minimum gas softening length of 0.074 comoving kpc. These resolutions enable accurate reproduction of kinematic properties for galaxies with $M_\star \geq 10^9 \,\mathrm{M_\odot}$ \citep{pillepich2019}.

We use friends-of-friends (FOF) \citep{Davis1985} and SUBFIND \citep{Springel2001} to identify halos and subhalos. Galaxies are defined as gravitationally bound particles within their host subhalos (stars, gas, dark matter, black holes). Each galaxy is oriented face-on based on stellar angular momentum and centered at the gravitational potential minimum, with bulk velocity subtracted.

\subsection{Sample selection}
We study disk galaxy profile types using a main sample of disk galaxies selected from the TNG50 simulation \citep{pillepich2019}. Such a sample is selected via three criteria: 
inclusion of only central galaxies to minimize environmental effects; a stellar mass cutoff $M_\star > 10^{10}\,M_{\odot}$, where $M_\star$ is computed from all stellar particles bound to the subhalo by SUBFIND, ensuring reliable kinematic property reproduction; and a disk mass fraction requirement $f_{\rm disk} > 0.4$ within three half-mass effective radii to guarantee prominent disk structures. Our main sample meeting all criteria comprises 423 galaxies, with their stellar mass distribution shown as the solid-black histogram in Figure \ref{fig:1}’s upper-right panel. The disk mass fraction $f_{\rm disk}$ is calculated via the dynamical decomposition method \citep[see details in][]{du2019, du2020}\footnote{The data of kinematic structures in TNG galaxies are publicly accessible at \url{https://www.tng-project.org/data/docs/specifications/\#sec5m}}, which employs automatic Gaussian Mixture Models (called \texttt{auto-GMM}) to cluster stellar particles in the kinematic phase space of circularity ($\varepsilon = j_z/j_c$, where $j_z$ is the stellar angular momentum along the rotation axis and $j_c$ is that of a circular orbit at the same binding energy), binding energy, and non-azimuthal angular momentum \citep{abadi2003, domenech2012}. Components with mean circularity $\langle j_z/j_c \rangle > 0.5$ are classified as disks: cold disks ($\langle j_z/j_c \rangle > 0.85$) and warm disks ($0.5 \leq \langle j_z/j_c \rangle \leq 0.85$).
%which employs automatic Gaussian Mixture Models (called \texttt{auto-GMM}) to cluster stellar particles in the kinematic phase space of circularity, binding energy, and non-azimuthal angular momentum \citep{abadi2003, domenech2012}.
This technique isolates a kinematically distinct disk component from the composite spheroid (encompassing both bulges and halos). The resulting dynamics-based decomposition is fundamental for analyzing disk profiles, as it enables us to probe how disks and spheroids differentially shape profile types. Our $f_{\rm disk} > 0.4$ threshold aligns with kinematically-defined disk galaxies and fast rotators, as suggested by \citet{zhong2026}. Such a criterion corresponds approximately to a cylindrical rotation energy significance of $\kappa_{\rm rot} > 0.5$ \citep{sales2010}—a widely-used criterion for identifying simulated disk galaxies. 

To enable robust comparison with observations, we supplement our analysis with a sample of 278 visually-classified disk galaxies with $M_\star > 10^{10}\,M_\odot$ (both centrals and satellites; dashed histogram in the upper-right panel of Figure~\ref{fig:1}). As shown in \citet{HeDu2025}, 30–60\% of morphologically disky systems host dominant spheroidal components \citep[e.g., Sombrero galaxy in][]{gadotti2012}. Such disks, formed by reformation after dry major mergers, are typically embedded in massive stellar halos \citep{du2021}. Despite exhibiting low kinematically-defined disk mass fractions, they retain observational disk classifications due to features like spiral arms, ongoing star formation, or bars.

\subsection{Data extraction and classification of surface density profiles}

We derived one-dimensional (1D) surface density profiles from the face-on view using radial annuli with a constant interval of 0.1 kpc. These profiles were fitted with a model consisting of a S\'ersic bulge and an exponential disk (see the left panels in Figure~\ref{fig:1} for examples), where the initial parameters were estimated from separate fits to the inner and outer regions under visual judgment. This fitting procedure yielded the disk scale length (\(h_R\)) and the bulge S\'ersic index (\(n\)). The outer boundary for the fit is set at the smaller of \(30\) kpc or the radius where the surface density falls below \(10^6 \,\mathrm{M_\odot\,kpc^{-2}}\), following \citet{chamba2020} and \citet{trujillo2020}. As illustrated in the left panels of Figure~\ref{fig:1}, we classify such galaxies based on the residuals of this fit in their outer parts: Type~I (residuals within \(\pm 0.2\) dex), Type~II (down-bending, residual \(< -0.2\) dex), and Type~III (up-bending, residual \(> 0.2\) dex). The break radius \(R_{\rm break}\), marked by black dashed lines, is defined as the largest radius where the residual is equal to zero.

Five galaxies from the parent sample were excluded due to ambiguous profiles. Our main sample comprises 418 central, disk-dominated galaxies, with subsamples classified as 161 Type~I (38.5\%), 141 Type~II (33.7\%), and 116 Type~III (27.8\%) disk profiles. As illustrated in the right-lower panel of Figure \ref{fig:1} (solid profiles and histograms), these types exhibit distinct stellar mass dependencies. Galaxies with a Type~I disk (yellow) account for $\sim30-50$\% of disk galaxies across a broad mass range ($10^{10}$–$10^{11.2}M_{\odot}$). Galaxies with a Type~II disk (blue) account for $\sim40$\% of disk galaxies at the low-mass end ($<10^{10.6}\,M_{\odot}$) but decline sharply at higher masses ($>10^{10.6}\,M_{\odot}$). Conversely, the Type~III (pink) fraction dramatically increases from 20\% to 60\% across the same mass range. We emphasize that our main sample, limited to kinematically-selected central disks, cannot directly benchmark morphology-based observational samples.

%fig2
\begin{figure*}[!t] % !t表示优先放在页面顶部 跨栏图片
\centering
\includegraphics[width=\textwidth]{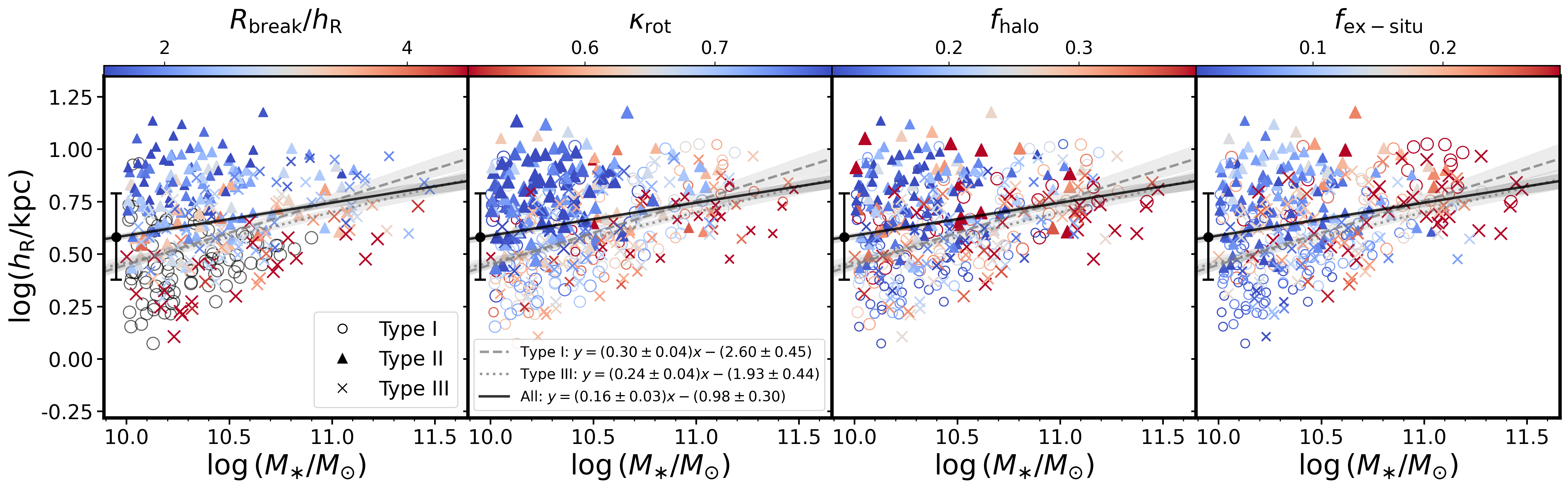}
\caption{
Kinematic and structural properties of Type~I (circles), Type~II (triangles), and Type~III (crosses) galaxies in the $h_R$--$M_\star$ plane. From left to right, the data points are color-coded by the ratio of break radius to inner disk scale length $R_{\rm break}/h_R$, rotational support $\kappa_{\rm rot}$, stellar halo mass fraction $f_{\rm halo}$, and ex-situ stellar mass fraction $f_{\rm ex\text{-}situ}$. In each panel, the symbol size scales with the corresponding color-coded quantity. The gray dashed, gray dotted, and black solid lines represent the linear fits to the $h_R$--$M_\star$ relation for Type~I, Type~III, and the total disk sample, respectively; shaded regions denote the 1$\sigma$ fitting uncertainties. The error bars indicate the standard deviation of the residuals from the global linear fit.}
\label{fig:2}
\end{figure*}

For direct comparison with observations, we combine our primary selection with visually-defined disk galaxies (including satellites) to create an augmented sample of 696 galaxies, as illustrated in the right panels of Figure \ref{fig:1} (dashed profiles and histograms). This inclusion of morphologically-selected galaxies does not significantly alter our findings in comparison with the main sample (solid profiles). 
Crucially, the augmented sample successfully reproduces both the mass-dependent trends and the relative abundances of the dominant profile types found by \citet{tang2020} (squares), who used SDSS $r'$-band images of nearly face-on disk galaxies in the redshift range $0.01 < z < 0.15$, with radial coverage extending to $\sim 2.5\,R_e$. We also compare our results with those of \citet{laine2016} (triangles), who classified 753 nearly face-on nearby galaxies with Hubble types $-2 < T \leq 9$ and $8.5 \leq \log_{10}(M_\star/M_\odot) \leq 11$, using $3.6\,\mu$m images from the S$^4$G survey and $K_s$-band images from the NIRSOS survey. Overall, the Type~II and Type~III fractions in TNG show broad agreement with observations: at low masses ($<10^{10.6}\,M_{\odot}$), Type~II disks dominate ($\sim 40$\% of galaxies), consistent with both \citet{tang2020} and \citet{laine2016}, while Type~III fractions show an increasing trend with stellar mass in both simulations and observations. However, \citet{laine2016} found Type~II fractions that increase with stellar mass, in contrast to the declining trend seen in both TNG and \citet{tang2020}. Notable discrepancies also remain for the Type~I fractions, which are systematically higher in our classification. This may be due to differences in classification criteria adopted across studies. At low masses, the discrepancy appears primarily between the Type~I and Type~III fractions, whereas at high masses it shifts to the Type~I and Type~II fractions. Notably, this discrepancy widens when visually classified disk galaxies are included, suggesting that sample selection methods are not the primary driver of these differences. The low Type~II fraction in the $10^{10.6}$--$10^{11}\,M_\odot$ range may partly arise from the limited volume of TNG-50, which undersamples massive galaxies, the efficient AGN kinetic feedback \citep{weinberger2018} that suppresses the young stars essential for prominent breaks, and the underrepresentation of bar-driven resonance effects \citep{munozmateos2013, laine2014}.

In the following analysis, we primarily focus on the main sample for simplicity. This study gains insights into the origins of the diverse surface density profiles of disk galaxies.

\section{General properties and merger impact of galaxies exhibiting different types of breaks}\label{sec:3}

\subsection{Difference in galaxy compactness and break radius}

Type~I (circles) and Type~III disks (crosses) exhibit fundamentally distinct scaling relations compared to Type~II cases (triangles), as shown in the mass–size diagram (Figure~\ref{fig:2}). Both Type~I and Type~III disk galaxies obey a mass–size relation characterized by greater compactness across a broad mass range ($> 10^{10} M_\odot$) – with Type~I/III disks dominating the high-mass regime ($ \gtrsim 10^{10.6} M_\odot$). In contrast, Type~II disks are exclusively confined to lower masses ($\lesssim 10^{10.6}M_\odot$) while showing anomalously large sizes. 
Linear fitting gives $\log (h_R/{\rm kpc}) = (0.30 \pm 0.04)\,\log (M_\star/M_\odot) - (2.60 \pm 0.45)$ for Type~I disks (gray dashed lines) and $\log (h_R/{\rm kpc}) = (0.24 \pm 0.04)\,\log (M_\star/M_\odot) - (1.93 \pm 0.44)$ for Type~III disks (gray dotted lines). The two relations have similar slopes and differ only slightly in $\log h_R$ at fixed stellar mass. For comparison, the fitting result of all disk galaxies is $\log (h_R/{\rm kpc}) = (0.16 \pm 0.03)\,\log (M_\star/M_\odot) - (0.98 \pm 0.30)$ (black solid lines). Most of the Type~II disk galaxies are above this line. 
Our findings align with observational results from \citet{wang2018}, who analyzed a comparable sample of face-on disk galaxies selected from SDSS $r'$-band imaging. In lower-mass galaxies with $\log(M_\star/M_\odot) \lesssim 10.6$, they demonstrated that the fraction of cases with a Type~II break increases sharply toward lower concentrations ($R_{90}/R_{50} < 2.2$), reaching values near 70\%. This correlation indicates that galaxies with Type~II breaks possess systematically more extended disks.

\begin{figure*}[!t]
\centering
\includegraphics[width=\textwidth]{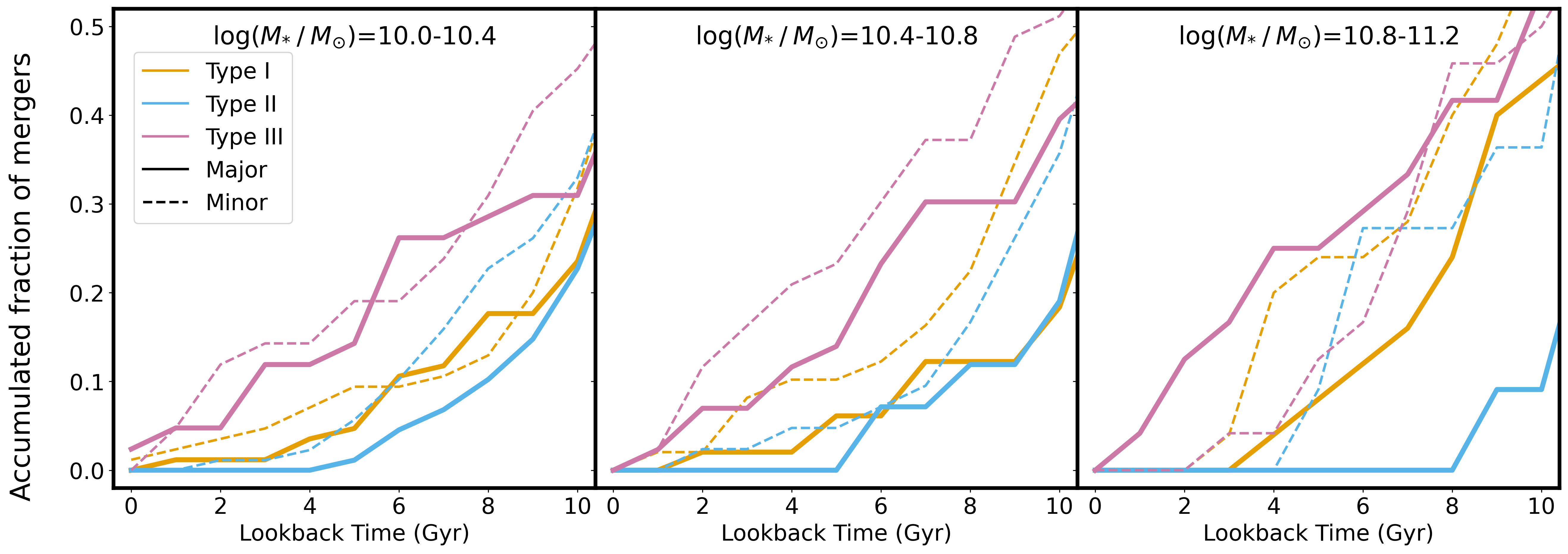}
\caption{Accumulated fraction of mergers for Type~I (yellow), Type~II (blue), and Type~III (pink) disk galaxies across three stellar mass bins: $\log(M_\star/M_\odot) = 10.0$--10.4 (left), 10.4--10.8 (middle), and 10.8--11.2 (right). The fraction of galaxies experiencing at least one major merger (solid profiles, mass ratio $> 1:4$) or minor merger (dashed profiles, mass ratio $1:4$--$1:10$) as a function of lookback time.}
\label{fig:3}
\end{figure*}

As evidenced in the first panel of Figure~\ref{fig:2}, Type~III profiles systematically exhibit larger normalized break radii ($R_{\rm break}/h_R$) compared to Type~II systems. We establish that this distinction stems primarily from differences in scale length $h_R$ rather than absolute break radii $R_{\rm {break}}$, which show comparable magnitudes between types. Type~II disks display larger $h_R$ (median = 6.39 {\rm kpc} , mean = 6.55 {\rm kpc}), resulting in somewhat smaller $R_{\rm break}/h_R$ (median = 2.21, mean = 2.29), while Type~III systems possess shorter $h_R$ (median = 4.09 {\rm kpc} , mean = 4.38 {\rm kpc}), producing extended normalized breaks  (median = 2.74, mean = 3.09). This parameter shows only weak stellar-mass dependence across our sample. Our findings align with \citet{pohlen2006}'s SDSS $r'$-band analysis reporting a smaller mean $R_{\rm break}/h_R \simeq$ 2.5 for Type~II versus $\simeq$4.9 for Type~III disks. TNG50 simulations thus plausibly reproduce observed distinctions between down-bending (Type~II) and up-bending (Type~III) profiles.

\subsection{Rotation and potential impact of mergers}
%krot
Kinematic measurements from Figure ~\ref{fig:2} (second panel) systematically differentiate the three disk types: Type~II disks exhibit the strongest rotational support (median $\kappa_{\rm rot}=0.75$), whereas Type~III disks show weaker rotational coherence (median $\kappa_{\rm rot}=0.63$). Type~I systems are kinematically intermediate (median $\kappa_{\rm rot}=0.66$). This result qualitatively aligns with \citet{wang2018}, who reported Type~II disk galaxies as exhibiting markedly higher angular momentum than their Type~I and Type~III counterparts. A comparable correlation between down-bending profiles and enhanced rotational support is also achieved in the EAGLE simulation, where \citet{pfeffer2022} report a steep rise in Type~II fraction with increasing $\kappa_{\rm co}$, reaching $\gtrsim 80\%$ in the highest-$\kappa_{\rm co}$ bin for field galaxies.

In the third and fourth panels of Figure~\ref{fig:2}, the stellar halo mass fractions ($f_{\rm halo}$) and the ex-situ stellar mass fractions ($f_{\rm ex\mbox{-}situ}$), both used as merger diagnostics, reveal a consistent ordering in merger activity. In-situ and ex-situ stars are distinguished based on their birth radius in spherical coordinates ($r_{\rm birth,3D}$): in-situ stars are those formed with $r_{\rm birth, 3D} \leq 30$ kpc, while ex-situ stars are those with $r_{\rm birth, 3D} > 30$ kpc.
We derive stellar birth coordinates by transforming native TNG simulation formation positions into galaxy rest frames using \texttt{GalaxyPose} \citep{Lu2026GalaxyPose}. Each frame is reconstructed per snapshot using the central position, bulk stellar velocity, and angular momentum vector evolution of progenitor galaxies. Orbital trajectories and orientation functions achieve temporal continuity through cubic spline interpolation and spherical linear interpolation (slerp), respectively. The resulting positional uncertainty is less than 1 kpc, which is negligible for both the radial density profile analysis and computation of global galaxy parameters performed here.

Type~II disks have the weakest merger signatures (median $f_{\rm halo}=0.14$, $f_{\rm ex\mbox{-}situ}=0.05$), Type~III the strongest (median $f_{\rm halo}=0.26$, $f_{\rm ex\mbox{-}situ}=0.19$), and Type~I systems lie in between (median $f_{\rm halo}=0.20$, $f_{\rm ex\mbox{-}situ}=0.08$). \citet{ruiz2017} obtained a similar conclusion using Milky Way analogues in the RaDES cosmological simulations. Figure~\ref{fig:3} traces the cumulative merger histories for the three galaxy types across three stellar mass bins:  $\log(M_\star/M_\odot)$ = 10.0–10.4 (low), 10.4–10.8 (intermediate), and 10.8–11.2 (high). We define the merger mass ratio $\mu$ as $\mu \equiv M_{\star,\rm sec}^{\rm max}/M_{\star,\rm pri}$, where $M_{\star,\rm sec}^{\rm max}$ is the maximum historical stellar mass attained by the secondary progenitor, and $M_{\star,\rm pri}$ is the primary's stellar mass at the final snapshot where SUBFIND \citep{Springel2001} identifies the secondary as an independent subhalo. Type~II (blue) galaxies exhibit the lowest merger frequencies for both major ($\mu > 0.25$; solid) and minor mergers ($0.1 \leq \mu \leq 0.25$; dashed). Conversely, Type~III (pink) systems show the highest major merger fractions across all masses, increasing with stellar mass, while their minor-merger fractions peak at low-to-intermediate masses. Type~I (yellow) displays intermediate major merger frequencies but dominates in minor mergers at high masses, suggesting Types~I and III result from distinct external perturbation regimes.

These results indicate Type~II break galaxies evolve predominantly through internal (``secular'') processes, i.e., the natural evolutionary pathway, retaining high rotational velocities and large disk sizes due to minimal merger/tidal disruption. Consequently, they exhibit stronger rotational support and less massive stellar halos than Type~I/III systems. In contrast, Type~III (up-bending) breaks originate primarily from merger events that supply accreted stars while disrupting outer disk structure \citep{bekki1998, aguerri2001, eliche2006}. Both major and minor mergers \citep{borlaff2014} likely contribute to Type~I/III break formation. Type~II systems thus serve as optimal laboratories for probing merger-free secular evolution and constitute primary targets for studying undisturbed assembly histories. Detailed analysis of Type~III disk galaxies is beyond the scope of this paper.

\section{Origin of Type~II Disks and Their U-shaped Age Profiles}\label{sec:4}
The formation of Type~II disk breaks is primarily interpreted through two physical channels: radial migration and star formation truncation. The first involves the outward redistribution of stars driven by non-axisymmetric features \citep{debattista2006, herpich2017}, which can explain observed U-shaped stellar age profiles \citep{zheng2015, dale2015}. Alternatively, the second channel attributes the breaks to an abrupt radial change in star formation profiles \citep{pfeffer2022, tang2020}. However, as star formation ceases beyond the break radius in this scenario, it cannot naturally account for the presence of old stellar populations at large radii. To reconcile this, these models typically invoke outward radial migration to transport old stars into the outer disk \citep{roskar2008, tang2020} or an extended stellar halo contribution \citep{pfeffer2022}; without these contributions, star-formation-driven models struggle to explain the stellar distribution beyond the break. In the following analysis, we investigate what mechanism primarily drives the formation of Type~II profiles in TNG50.

The TNG simulations reasonably model gas cycles and star formation physics at low redshifts ($z \leq 1$), as evidenced by their consistency with SDSS observations of the galaxy stellar mass function (GSMF) and color bimodality \citep{nelson2018first}, and significantly improved agreement with HI data relative to its predecessor Illustris \citep{guo2020}. This robust validation establishes TNG as a credible theoretical framework for studying structural evolution and formation mechanisms of Type~II disk galaxies.

\subsection{A New Channel Driven by Inside-out Growth of Cold Gas Accretion and sSFR}

\begin{figure}[!t]
\centering
\includegraphics[width=0.95\columnwidth]{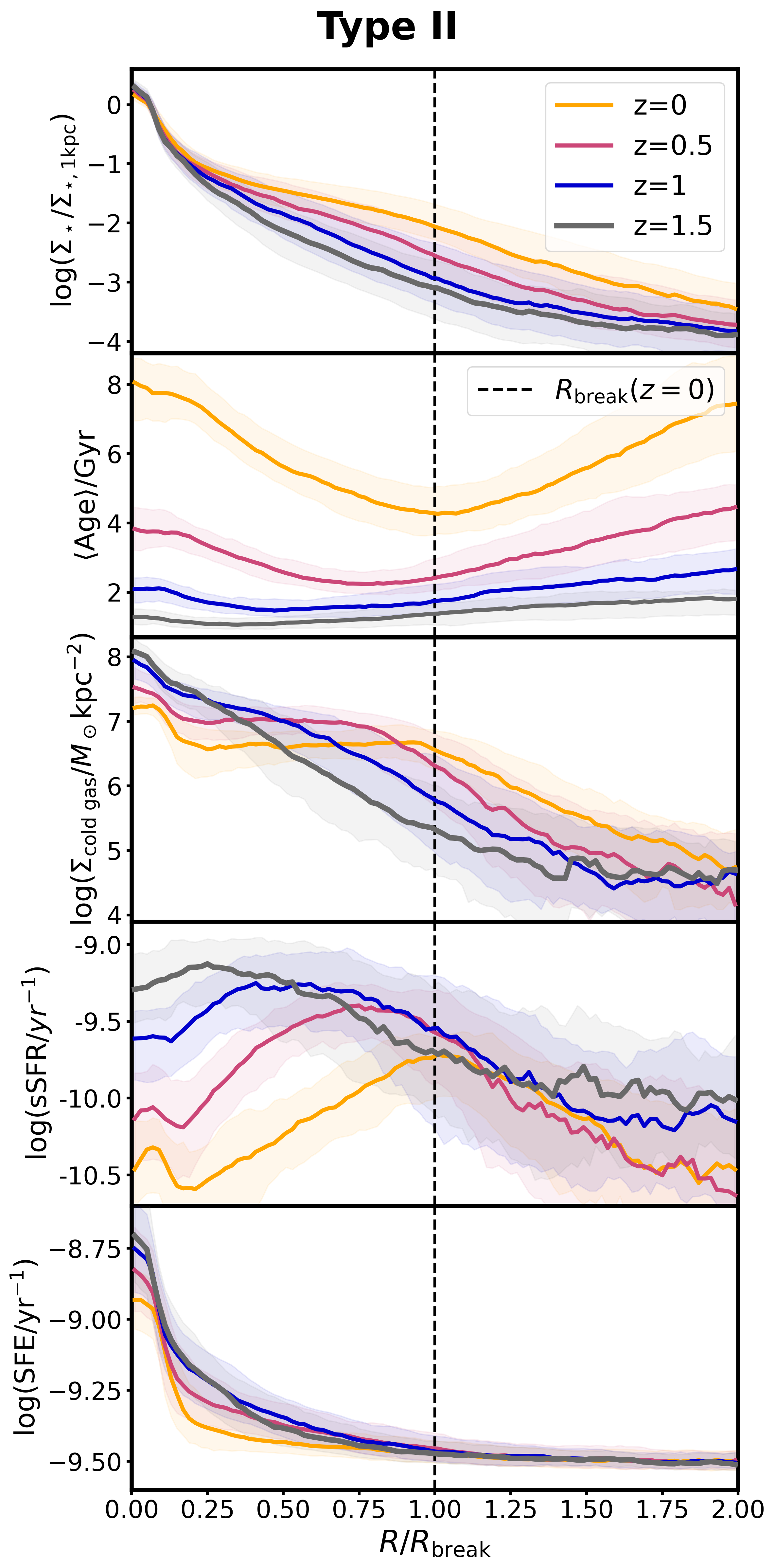}
\caption{Evolution of radial profiles for Type~II disk galaxies from $z=1.5$ to $z=0$, plotted against the normalized radius $R/R_{\rm break}$. The panels represent, from top to bottom: normalized stellar surface density ($\Sigma_\star / \Sigma_{\star, 1\rm{kpc}}$), mass-weighted mean stellar age computed from all stellar particles within a cylindrical radius of $R \leq 30$~kpc, cold gas surface density ($\Sigma_{\rm cold\,gas}$), specific star formation rate (sSFR), and star formation efficiency (SFE). Colored curves correspond to different redshifts: $z=0$ (yellow), $z=0.5$ (pink), $z=1.0$ (blue), and $z=1.5$ (gray). Solid lines denote the median values, while shaded regions indicate the interquartile range (25th–75th percentiles). The vertical dashed line marks the break radius identified at $z=0$.}
\label{fig:4}
\end{figure}

To reveal the physical drivers behind the formation of Type~II disks, we trace the evolution of their radial profiles from $z=1.5$ to $z=0$ (Figure~\ref{fig:4}). The solid profiles show the median values for Type~II disk galaxies as a function of normalized radius ($R/R_{\rm break}$), where each galaxy profile was first computed in $0.3\,\mathrm{kpc}$ annuli and then mapped to normalized radius using $R_{\rm break}$ ($z=0$). For consistency, $\Sigma_{\rm cold\,gas}$ is computed using gas cells selected to lie close to the disk mid-plane ($|z|\le 3\,{\rm kpc}$) and to have ${\rm SFR}>0$ within a cylindrical radius of $R \le 30\,{\rm kpc}$; the same selection is applied when computing $\Sigma_{\rm SFR}$ (and hence sSFR and SFE). The stellar age profile calculation includes all stars within a cylindrical radius of $R \le 30\,{\rm kpc}$. We find that the profiles exhibit synchronized evolutionary features. At $z \geq 1$ (blue and gray), the normalized galaxy density profiles generally have no down-bending break and the age profiles are quite flat, as shown in the first and second panels of Figure~\ref{fig:4}, with no U-shape present. As the system evolves to $z \sim 0.5$ (pink), a Type~II break in the normalized density profile gradually emerges, closely accompanied by the formation of a U-shaped age profile. This feature persists and strengthens until $z=0$ (yellow). A flattened U-shaped age profile in older stellar populations is also found in TNG Type~II galaxies, consistent with the findings of \citet{radburn2012} (see Appendix~\ref{app:A} for details).

This structural transformation coincides closely with the evolution of the cold-gas distribution and the specific star formation rate (sSFR), as shown in the third and fourth panels of Figure~\ref{fig:4}. The stellar surface-density profile at $z \sim 1$ remains nearly single-exponential beyond the present-day break radius; consequently, the secular flattening of $\Sigma_{\rm cold,gas}$ since $z \leq 0.5$ (pink and yellow) induces a localized enhancement of sSFR around $R_{\rm break}$, thereby producing a radial break. This localized enhancement drives increased stellar mass assembly at $R_{\rm break}$ relative to both the interior and exterior disk regions, resulting in a down-bending exponential profile. A similar connection between $R_{\rm break}$ and the gas/star-formation rate (SFR) profile truncation was also noted by \citet{roskar2008}, aligning with our findings in TNG50. Importantly, the bottom panel of Figure~\ref{fig:4} confirms that the star formation efficiency (SFE) profiles remain spatially flat and temporally constant. This demonstrates that the radial break does not originate from localized variations in star formation physics, but is a structural consequence of the sSFR peak. Moreover, the truncation radius of the cold gas surface density ($\Sigma_{\rm cold\,gas}$), the peak position of the sSFR, as well as $R_{\rm break}$ and the minimum of the U-shaped age profile, are consistent 
and all shift outward with time. This effectively explains the phenomenon that the radial position of the break has been found to increase with cosmic time \citep{trujillo2005, azzollini2008}.

\begin{figure}[!t]
\centering
\includegraphics[width=0.9\columnwidth]{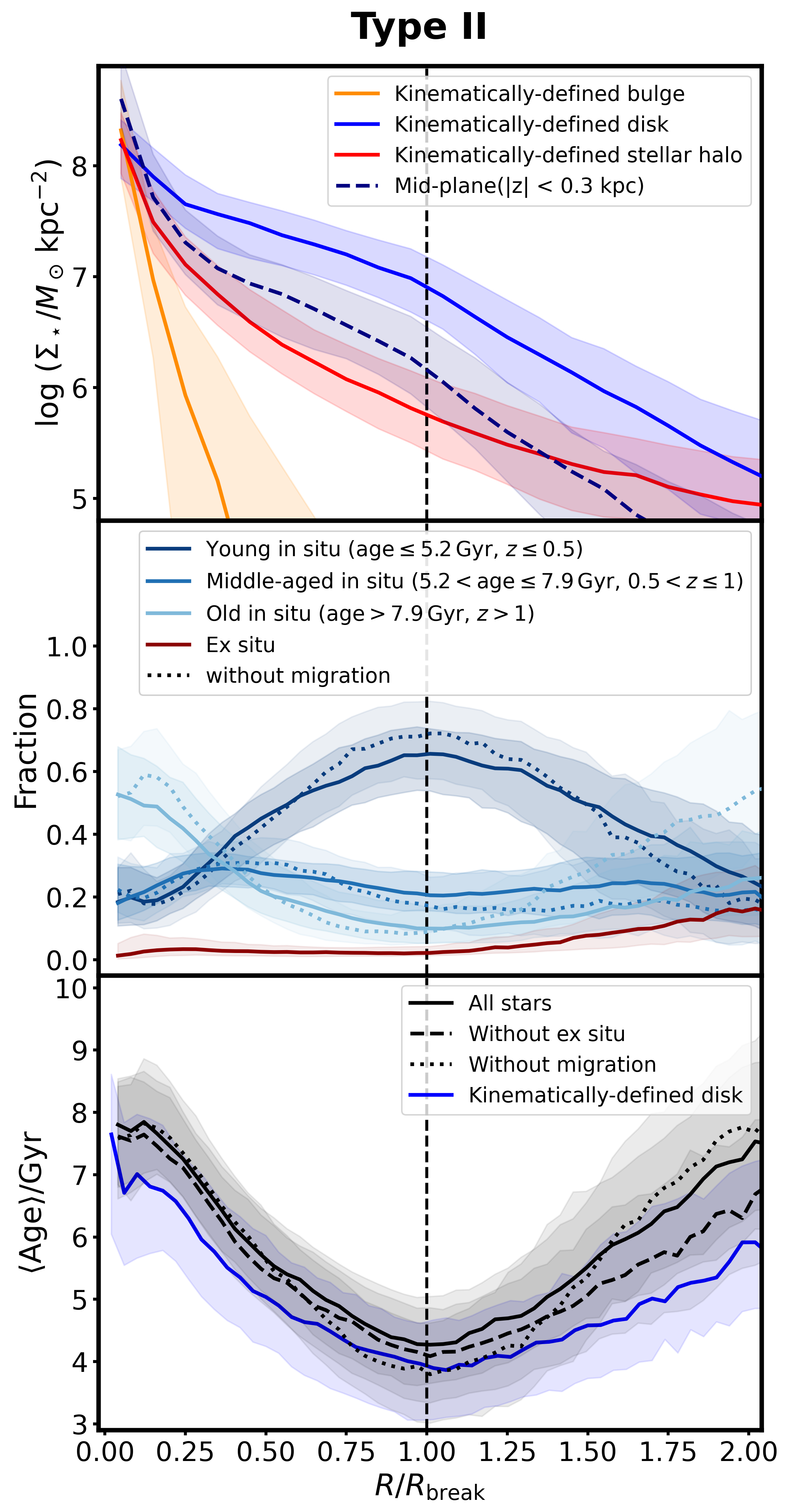}
\caption{
\textbf{Top panel:} radial surface density profiles for the kinematically-defined disk (blue), bulge (yellow), and stellar halo (red), following the decomposition method of \citet{du2019, du2020}, along with the surface density profile of mid-plane stars ($|z| < 0.3$~kpc, dark blue dashed).
\textbf{Middle panel:} radial mass fractions of ex-situ stars (brown) and in-situ stars. The in-situ population is subdivided into three age bins: young, intermediate-age, and old (dark to light blue shading). Dotted lines represent the distributions excluding radial migration.
\textbf{Bottom panel:} radial mass-weighted age profiles for all stellar particles within a cylindrical radius of $R \leq 30$~kpc (black) and the corresponding distributions after excluding ex-situ stars and radial migration, along with the age profile of the kinematically-defined disk component (blue).
In each panel, solid lines denote median values, shaded regions indicate the interquartile range (25th--75th percentiles), and the vertical dashed line marks the break radius $R_{\rm break}$.
}
\label{fig:5}
\end{figure}

While \citet{sanchez2009} similarly observed a Type~II break in the cosmological hydrodynamical simulation of a single disk galaxy, these features manifested exclusively in young stellar populations, with no discernible break in the total surface density profile. They attributed the U-shaped age profiles primarily to the sharp SFR decline beyond the break radius—a mechanism related to but distinct from our results, as their sharp decline in SFR happened in warps. Moreover, we identify localized sSFR enhancement as the direct driver simultaneously producing both Type~II breaks and U-shaped age profiles. Since sSFR quantifies the relative formation rate of young stars across radii, it provides the fundamental link between star formation dynamics and structural evolution. Furthermore, we visually verified that inside-out growth of sSFR peaks has no clear link with the appearance of warps.

The Type~II break constitutes a fundamental structural signature of dynamical cold disks themselves, rather than a feature that merely reflects the relative importance of the disk compared to the spheroid. By utilizing the kinematic decomposition from \citet{du2019, du2020}, the top panel of Figure~\ref{fig:5} shows clearly that the outer regions (extending to $2 \times R_{\text{break}}$) are dominated by rotationally supported disk components. The break radius identified in the kinematic disk density profile aligns with that of the total surface density profile. We further verified that mid-plane stars ($|z| < 0.3\,\text{kpc}$) of Type~II disk galaxies exhibit a prominent down-bending break in their surface density profile (dark-blue profile in the top panel); furthermore, the cold kinematics of the outer disc confirm that the age upturn beyond $R_{\rm break}$ is free from spheroidal contamination (see Appendix~\ref{app:B} for details). These results suggest that the formation of Type~II breaks is largely dominated by rotationally supported structures rather than spheroidal components (e.g., the stellar halo). For Type~II disk galaxies in their comparable mass range ($\log(M_\star/M_\odot) \sim 10–11$), stellar halos show systematically lower surface densities of $\sim1$ dex than their disk counterparts, despite minor differences in total stellar mass among galaxies (which we neglect). Such stellar halos thus exhibit negligible contributions beyond $R_{\rm break}$ and cannot reproduce the outer-disk part of the observed U-shaped age profiles.

Type~II profile formation aligns closely with the ``extended galaxies'' evolutionary track identified in \citet{ma2024}. The emergence of Type~II features at $z \sim 0.5$ stems from delayed angular momentum ($j$) transfer: dark matter halo $\rightarrow$ total gas $\rightarrow$ cold gas. This process yields a few Gyr lag between the specific angular momentum evolution of cold gas ($j_{\rm cold\ gas}$, shown in their Figure 6) and total gas ($j_{\rm gas}$). The delay implies that cosmic environment-driven $j$-gains require finite time to restructure cold gas distributions and stellar disks. Physically, late-phase accretion ($z < 1$) injects high-$j$ gas that settles preferentially at outer regions of disks rather than their centers, directly driving the inside-out growth of cold gas density and Type~II breaks.

\subsection{Demographics of Stars of Down-bending Breaks: In-situ Formation, Radial Migration, and Ex-situ Accretion}

Building on evidence from the preceding section that Type~II breaks form gradually at $z < 1$, we isolate potential contributions from alternative mechanisms by decomposing $z = 0$ stellar populations by origin (in situ vs. ex situ) and formation time. 
Further, we divide in-situ stars into three age cohorts shown in Figure~\ref{fig:5} (middle panel, from dark to light blue): young (age $\leq 5.2$ Gyr, $z \leq 0.5$), middle-aged (5.2 Gyr $<$ age $\leq 7.9$ Gyr, $0.5 < z \leq 1$), and old (age $>$ 7.9 Gyr, $z > 1$). 
Note that this decomposition includes all stars in the galaxy rather than being restricted to kinematically-defined disk components.
In Type~II systems, young in-situ stars dominate the stellar mass budget around $R_{\rm break}$, contributing $\sim 70$\% of the total, while ex-situ stars remain negligible ($<10$\%, brown profile) across a broad radial range, extending out to $\sim 1.7\,R_{\rm break}$, and thus cannot explain U-shaped age profiles. 
To isolate radial redistribution effects, we compute the birth-radius-dependent fraction of young stars (dark-blue dotted lines) using the cylindrical birth radius ($R_{\rm birth, 2D}$), representing a no-migration scenario.
Although older stars migrate over longer timescales, the comparison between profiles with and without migration (solid vs. dotted lines) shows that they diffuse throughout the disk rather than accumulating in the outskirts, leading to a decreased relative fraction in the outer disk. As demonstrated in the bottom panel of Figure~\ref{fig:5}, the age profile of the kinematically defined disk component exhibits a clear U-shaped profile (blue line), confirming that the age upturn is not an artifact of spheroidal contamination. Our comparative analysis further shows that removing ex-situ stars has a minimal effect on the age gradient (black-dashed line), while excluding the migration effect actually yields a stronger and steeper U-shaped upturn (black dotted line). These findings confirm that radial migration is not the cause of the U-shape; instead, it acts to weaken both the break \citep{sanchez2009, tang2020} and the age upturn by diluting the localized stellar assembly at $R_{\rm break}$.
%As demonstrated in the bottom panel of Figure~\ref{fig:5}, our comparative analysis shows that removing ex-situ stars has a minimal effect on the age gradient, while excluding the migration effect actually yields a stronger and steeper U-shaped upturn. These findings confirm that radial migration is not the cause of the U-shape; instead, it acts to weaken both the break \citep{sanchez2009, tang2020} and the age upturn by diluting the localized stellar assembly at $R_{\rm break}$.

For quantitative analysis, measurements for Type~II disk galaxies reveal the origins of stars in the outer disk ($R > R_{\rm break}$). To account for orbital eccentricity of stars, migrated stars are specifically defined as those with $R_{\rm birth, 2D} < R_{\rm break} - 2\,\mathrm{kpc}$. Based on this criterion, median contributions to the outer disk comprise 60\% in-situ formation, 33\% migrated in-situ stars from inner regions, and 5\% ex-situ stars from early mergers, all of which are calculated using stars across all age bins.

\citet{lu2025} showed that in TNG simulations, both bar fractions and lengths are smaller than observed galaxies with $M_\star < 10^{10.5}\,M_\odot$ — the mass regime where Type~II disks prevail. Although this may reduce radial migration efficiency, we emphasize that 45\% of Type~II disks remain unbarred. The high prevalence of Type~II breaks in unbarred galaxies suggests that bars play a subordinate role in Type~II break formation.

Beyond the local Universe, we also compare our classification with high-redshift observations from \citet{xu2024,yu2025}, who classified 247 nearly face-on disk galaxies with $\log(M_\star/M_\odot) > 10$ at $1 \leq z \leq 3$ using JWST/CEERS F356W images. For a comparable sample of 247 disk galaxies at $z=1$ in TNG-50, we find Type~I: 36.1\%, Type~II: 9.5\%, and Type~III: 54.4\%, whereas \citet{xu2024} reported Type~I: 12.6\%, Type~II: 56.7\%, and Type~III: 34.8\%. The substantially lower Type~II fraction lies beyond this study's scope, given the challenges in both simulating and observing high-redshift galaxies. Moreover, the U-shaped age profile is not prominent in high-redshift Type~II galaxies, consistent with the observations of \citet{yu2025}. At high redshift, galaxies are too young to have developed a pronounced age contrast between the inner and outer disc, such that the amplitude of the U-shaped signal may not yet have reached a level that can be clearly identified, consistent with both high-redshift observations \citep[e.g.,][]{yu2025} and TNG simulations. The radial profile evolution of high-redshift Type~II galaxies is considerably more complex, and a detailed investigation of the underlying mechanisms is beyond the scope of this paper. We note that inside-out growth is the primary formation mechanism for Type~II breaks in our sample, while the influence of other mechanisms cannot be ruled out, particularly at high redshift where observational uncertainties remain large.
%Overall, these high-redshift comparisons do not affect our conclusions, which are based on the evolutionary histories and properties of Type~II disk galaxies identified at $z=0$.

\section{Summary}\label{sec:5}

Our analysis of 418 central disk galaxies ($M_\star > 10^{10}\,M_\odot$) from the TNG50 simulation demonstrates that simulated galaxies reproduce key observed trends: mass-dependent abundances, structural scaling relations, and the morphological segregation where Type~II disks dominate low-mass systems ($\sim40\%$) with unusually large sizes, while Type~III becomes prevalent at high masses ($>10^{10.6}\,M_{\odot}$) with greater compactness.

Systematic diagnostics reveal fundamental differences in disk types: Type~II systems show signatures of dynamically cold, undisturbed disks – maximal rotational support, minimal spheroidal and ex-situ components (median $\sim14$\% and $\sim5$\% respectively), and quiescent merger histories. Type~III systems exhibit the opposite characteristics: weakest rotational coherence, highest spheroidal and ex-situ fractions (median $\sim26$\% and $\sim19\%$ respectively), and merger-driven evolution. Type~I consistently occupies intermediate parameter space. This supports a framework where Type~II profiles represent the intrinsic disk state in isolation, with Type~I and Type~III emerging from progressively stronger external perturbations – moderate and major disturbances, respectively.

A new scenario is proposed that the inside-out growth of cold-gas accretion and star formation causes Type-II breaks and their associated U-shaped age profiles. Tracking these features from $z=1.5$ to $z=0$ reveals a critical transition around $z\sim0.5$, when both properties develop concurrently with: (1) long-term flattening of the cold gas density profiles, and (2) a localized sSFR peak around the break radius. This concentrated starburst accelerates stellar mass assembly, preferentially flattening the inner surface density profile relative to the outer disk. Such a break gradually expands to larger regions. Beyond the Type-II break radius, $\sim33$\% of stars originate from radial migration of stars born at inner regions, and only $\sim5$\% of stars are accreted from ex-situ origin. Galaxies with a Type~II break are thus likely to be ideal cases for studying the evolution of galaxies in nature where both mergers and environmental effects are minimal.

\begin{acknowledgments}
We are grateful to the referee for the insightful and constructive report, which has substantially improved the quality and clarity of the letter. This work is supported by the National Key R\&D Program of China (No. XY-2025-1459), the National Natural Science Foundation of China under grant No. 12573010, and the Science Fund for Creative Research Groups of the National Natural Science Foundation of China (No. 12221003). J.L. would like to acknowledge the NSFC under grant 12273027. LCH is supported by the National Science Foundation of China (12233001) and the China Manned Space Program (CMS-CSST-2025-A09). The TNG50 simulation used in this work, one of the flagship runs of the IllustrisTNG project, has been run on the HazelHen Cray XC40-system at the High Performance Computing Center Stuttgart as part of project GCS-ILLU of the Gauss Centers for Supercomputing (GCS). This work is also strongly supported by the Computing Center in Xi’an, China.
\end{acknowledgments}

\bibliography{sample701}{}
\bibliographystyle{aasjournalv7}

\appendix

This appendix presents two supplementary analyses in support of the mechanism proposed in this work: Appendix~\ref{app:A} validates well-known characteristic features of Type~II breaks, and Appendix~\ref{app:B} confirms the cold disc kinematics of Type~II disks.

\section{Properties of Type~II Disk Galaxies}\label{app:A}

\begin{figure}[!t]
\centering
\includegraphics[width=\columnwidth]{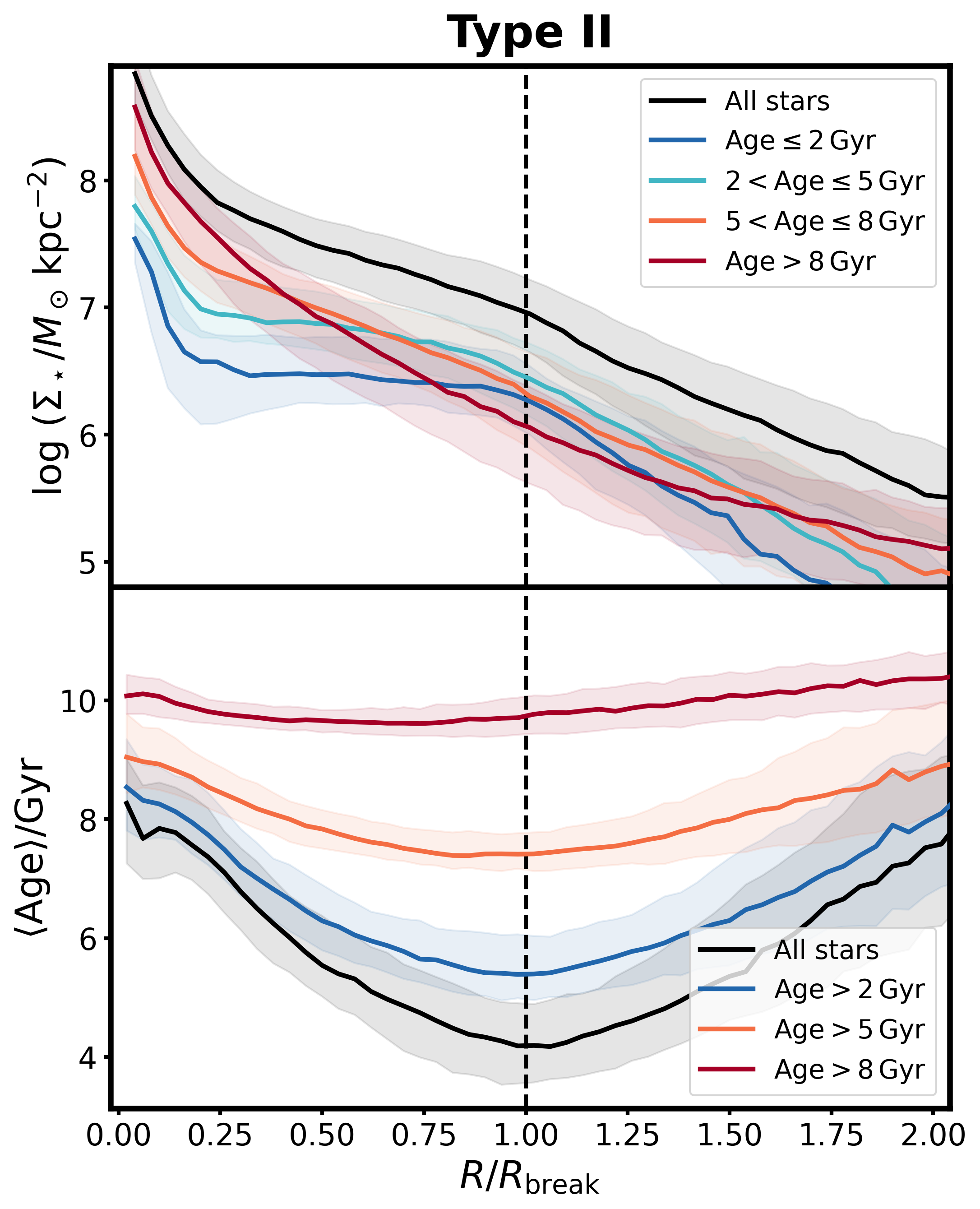}
\caption{
\textbf{Upper panel:} Median stellar surface density profiles of Type~II disk galaxies for different stellar age populations: all stars, $\leq$2~Gyr, 2--5~Gyr, 5--8~Gyr, and $>$8~Gyr.
\textbf{Lower panel:} Mass-weighted age profiles for progressively older stellar populations: all stars, $>$2~Gyr, $>$5~Gyr, and $>$8~Gyr.
}
\label{fig:6}
\end{figure}

\begin{figure}[!t]
\centering
\includegraphics[width=0.92\columnwidth]{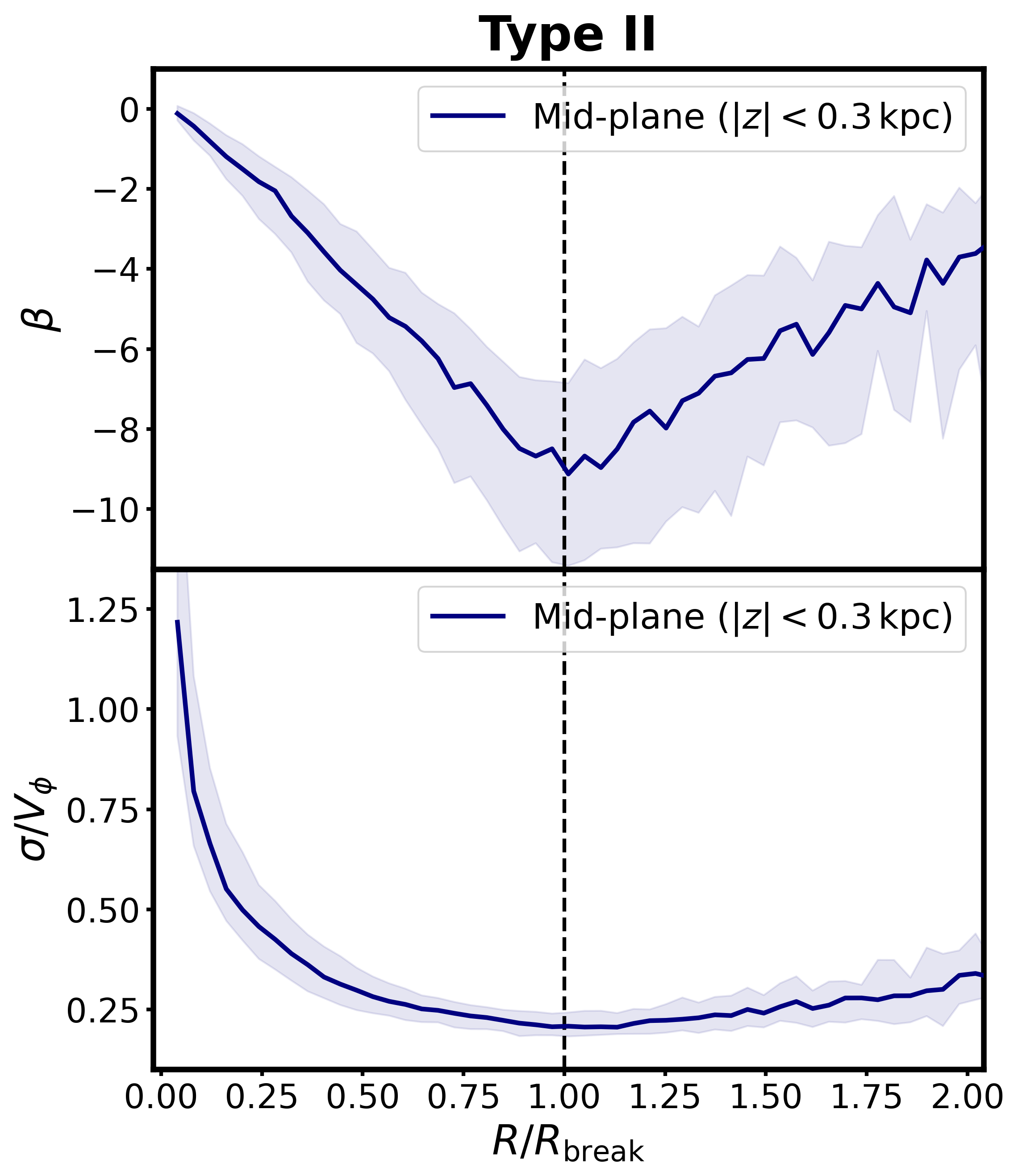}
\caption{
Kinematic diagnostics of Type~II disk galaxies. 
\textbf{Upper panel:} velocity anisotropy parameter $\beta$ as a function of normalized radius $R/R_{\rm break}$, computed from mid-plane stars ($|z| < 0.3$~kpc).
\textbf{Lower panel:} ratio of velocity dispersion to mean rotational velocity $\sigma/v_\phi$, computed from mid-plane stars ($|z| < 0.3$~kpc). 
}
\label{fig:7}
\end{figure}

We divide the stellar population into four age bins: $\leq$2~Gyr, 2--5~Gyr, 5--8~Gyr, and $>$8~Gyr. The upper panel of Figure~\ref{fig:6} shows the surface density profiles for each age bin. The youngest population (dark blue) exhibits the sharpest down-bending break, while the break weakens progressively with increasing stellar age. The oldest population (brown) shows a significantly flattened profile with no discernible break, consistent with the findings of \citet{radburn2012}. This is a direct consequence of the inside-out growth mechanism proposed in this work: the strongest break appears in the youngest population where the sSFR peaks. The lower panel of Figure~\ref{fig:6} shows the mass-weighted age profiles. The age profiles become increasingly flat with stellar age, and the oldest population (brown) exhibits an essentially flat radial age distribution. Consistent with inside-out growth of sSFR, Type~II breaks primarily trace younger stellar populations. This naturally explains their enhanced prominence in bluer photometric bands, where young stars dominate emission.

\section{Kinematic Properties of Type~II Disk Galaxies}\label{app:B}

To verify that the outer disk beyond $R_{\rm break}$ is kinematically cold and free from significant spheroidal contamination, we present kinematic diagnostics for Type~II disk galaxies in Figure~\ref{fig:7}, computed from all stellar particles within the mid-plane region ($|z| < 0.3$~kpc).
In the upper panel, the velocity anisotropy parameter $\beta = 1 - (\overline{v_\theta^2} + \overline{v_\phi^2}) / (2\overline{v_r^2})$, where $v_r$, $v_\theta$, and $v_\phi$ are the velocity components in polar coordinates, remains consistently negative across the entire radial range, indicating that stellar orbits are dominated by tangential motion. This holds even beyond $R_{\rm break}$, ruling out significant contamination from spheroidal or halo components whose orbits are radially biased. The inflection point of $\beta$ coincides with $R_{\rm break}$, reflecting the concentration of dynamically cold young stars at this location, consistent with our conclusions.
In the lower panel, the ratio $\sigma/v_\phi$, where $\sigma = \sqrt{(\sigma_R^2 + \sigma_\phi^2 + \sigma_z^2)/3}$ is the root-mean-square velocity dispersion of the three cylindrical components (radial, azimuthal, and vertical), and $v_\phi$ is the average cylindrical rotational velocity within each radial bin, decreases from $\sim$1.25 at the center to $\sim$0.25 by $0.5\,R_{\rm break}$ and remains at this low level out to $\sim$1.5\,$R_{\rm break}$ with only a marginal increase. This confirms that ordered rotation overwhelmingly dominates over random motion throughout the disk, including beyond $R_{\rm break}$.

\end{CJK*}
\end{document}